# *Effective Medium Response of Metallic Nanowire Arrays with a Kerr-type Dielectric Host*

*Mário G. Silveirinha*

*University of Coimbra, Department of Electrical Engineering – Instituto de Telecomunicações, Portugal, mario.silveirinha@co.it.pt*

**Abstract**

We derive an effective medium model to characterize the macroscopic electromagnetic response of metallic nanowire arrays embedded in a host dielectric with a Kerr-type nonlinear permittivity function. It is shown that the macroscopic electromagnetic fields are coupled to the conduction current in the nanowires and to an additional quasi-static potential through a system of nonlinear equations. We prove that a weak nonlinearity leads to an electromagnetic response closer to that of an indefinite medium, and to isofrequency contours with increased hyperbolicity. For high-field intensities the negative refraction of electromagnetic waves at an air − nanowire material interface is enhanced when the nanowires are embedded in a self-focusing Kerr medium.

# I. Introduction

Arrays of metallic nanowires stand as one of the most important structures in the metamaterial realm due to their applications in electromagnetic field manipulation and imaging in the nanoscale [1-3], negative refraction in the optical domain [4-7], in controlling the spontaneous emission of light by quantum-emitters [8, 9, 10], in enhancing the Cherenkov radiation by moving charges with no velocity threshold [11, 12], in providing a giant radiative heat transfer [13], and even in quantum electrodynamics in the framework of the Casimir effect, such that the Casimir interaction in a nanowire background can lead to ultra-long range forces [14, 15].

Nanowire materials can also lead to interesting physics in the context nonlinear optics, and in particular it was predicted in Ref. [16] that *subwavelength* solitons can be formed in arrays of metallic nanowires embedded in a Kerr medium, both in case of a self-focusing (staggered solitons) and self-defocusing media (unstaggered solitons). Earlier, it had been shown that analogous discrete solitons can be as well formed in layered metal-dielectric structures [17], and in general the formation of discrete solitons in arrays of coupled waveguides has also been topic of intense research [18, 19]. Several interesting features of discrete subwavelength lattice solitons in metallic nanowire arrays (e.g. the possibility of stable vortex and multipole solitons) have been investigated in recent works [20, 21, 22]. Such nonlinear modes can have an important impact in nanophotonics and in the realization of ultra-compact devices. Thus, it would be highly interesting to characterize them using an effective medium approach since this can greatly simplify the numerical modeling and highlight the relevant physics. As a first step in this direction, here, by extending the analytical model for the wire medium developed in previous works [23, 24, 25], we derive an effective medium theory that describes the dynamics of the macroscopic electromagnetic fields in a nanowire array embedded in a Kerr-type dielectric. Nonlinear effects in electromagnetic metamaterials have been investigated in many previous works, both the effective medium properties [26-28] as well as the general implications on the wave phenomena [29-31].

It is well known that at optical frequencies arrays of metallic nanowires behave as indefinite media, and that such property may lead to the negative refraction of waves at an interface with air [4-7]. The negative refraction is a consequence of the exotic hyperbolic-type dispersion of the photonic states in the metamaterial [4-5]. Here, we apply the proposed effective medium model to characterize the impact of the nonlinear effects on the negative refraction, showing that in case of a Kerr self-focusing medium the nonlinear response results in an enhancement of the negative refraction.

This paper is organized as follows. In Sect. II we generalize the theory of Ref. [25] to the case of a nonlinear host, showing that the dynamics of the macroscopic fields can be described with the help of two additional variables with known physical meaning. In Sect. III, we obtain the formulas for the (nonlinear) parameters of the effective medium model in terms of the macroscopic fields. Then, in Sect. IV we apply the developed theory to characterize the plane wave natural modes and investigate how the nonlinear response changes the isofrequency contours. In Sect. V, we study the negative refraction problem and compute the variation of the angle of transmission with the intensity of the transmitted energy density flux. Finally, in Sect. VI a conclusion is drawn.

## II. Geometry and Quasi-static Model

We consider an array of metallic nanowires embedded in a nonlinear host material with a Kerr-type nonlinearity such that for a fixed frequency the dielectric function may be modeled as:

$$\varepsilon_h = \varepsilon_h^0 \left(1 + \delta\varepsilon\right). \tag{1}$$

where $\delta\varepsilon$ is some nonlinear function of the complex electric field $\mathbf{e} = \mathbf{e}(\omega, \mathbf{r})$. In this work, we focus our attention in the case wherein the nonlinear response is the form $\delta\varepsilon = \alpha \mathbf{e}^* \cdot \mathbf{e}$, but the analysis can be readily extended to accommodate other more general microscopic responses. The parameter $\alpha$ may be estimated as $\alpha = 3\chi^{(3)}/\varepsilon_{h,r}^0$ being $\varepsilon_{h,r}^0 = \varepsilon_h^0/\varepsilon_0$ [32]. The nanowires are made of a metal with electrical permittivity $\varepsilon_m$, have radius $r_w$, and are arranged in a square lattice with lattice constant $a$ (Fig. 1). The nanowires are oriented along the $z$-direction, and the metal electrical response is assumed linear.

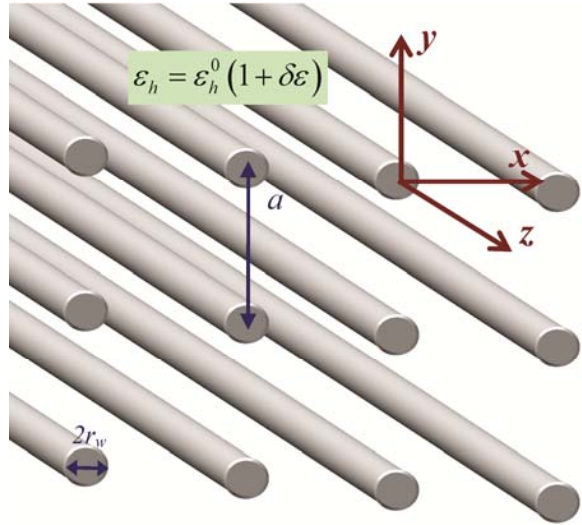

Fig. 1. (Color online) Geometry of the nanowire array. The lattice constant is $a$ and the radius of the metallic wires is $r_w$. The nanowires are embedded in a Kerr-type nonlinear host material with $\varepsilon_h = \varepsilon_h^0 (1 + \delta\varepsilon)$, being $\delta\varepsilon$ a function of the microscopic electric field.

The "microscopic" Maxwell's Equations (i.e. before any averaging on the scale of the structure period) are written in the frequency domain in the usual manner:

$$\nabla \times \mathbf{e} = +i\omega\mu_0 \mathbf{h}, \tag{2a}$$

$$\nabla \times \mathbf{h} = \mathbf{j}_{ext} - i\omega\varepsilon \mathbf{e}. \tag{2b}$$

Our objective is to obtain an effective medium description of the interaction of electromagnetic waves with the nanowires and the host material. This problem has been addressed previously in the literature for the case of *linear* materials [23, 24, 25]. The ideas described below are a generalization of the results of Ref. [25] to the case of a nonlinear host.

To begin with, we introduce an averaging operator $\langle \ \rangle$ so that the macroscopic electric field is related to the microscopic electric field by $\mathbf{E} = \langle \mathbf{e} \rangle$. Other macroscopic quantities are defined similarly. Note that the microscopic fields ($\mathbf{e}, \mathbf{h}$, etc) are denoted with lower-case letters, whereas the macroscopic fields ($\mathbf{E}, \mathbf{H}$, etc) are denoted with upper-case letters. The *fluctuating part* of the microscopic electric field is given by

$$\mathbf{e}_\Delta = \mathbf{e} - \mathbf{E}. \tag{3}$$

From this definition it follows that $\langle \mathbf{e}_\Delta \rangle = 0$, i.e. the result of averaging the fluctuating part of the electric field yields the null function.

It is supposed that $\langle \ \rangle$ represents a spatial convolution with some suitable test function, so that the form of Maxwell's equations is preserved after spatial averaging [33, 34]. In particular, the result of filtering a complex exponential of the form $e^{i\mathbf{k}\cdot\mathbf{r}}$ is always an exponential of the same type: $\langle e^{i\mathbf{k}\cdot\mathbf{r}}\rangle = F(\mathbf{k})e^{i\mathbf{k}\cdot\mathbf{r}}$ where $F(\mathbf{k})$ is some characteristic function. In this work, $\langle\ \rangle$ is taken as an ideal low-pass spatial filter, so that $F(\mathbf{k})=1$ when the wave vector $\mathbf{k}$ is within the first Brillouin zone of the metamaterial, and $F(\mathbf{k})=0$ otherwise. Finally, it is supposed that the external excitation, represented in (2) by the current density $\mathbf{j}_{ext}$, is inherently macroscopic. In other words, $\mathbf{j}_{ext}$ should stay invariant after spatial averaging: $\mathbf{j}_{ext} = \langle \mathbf{j}_{ext}\rangle$. This restriction implies that the external excitation cannot be more localized in space than the characteristic period of the metamaterial.

Within these hypotheses, it is easily found from (2) that the macroscopic fields satisfy:

$$\nabla \times \mathbf{E} = +i\omega\mu_0\mathbf{H}, \tag{4a}$$

$$\nabla \times \mathbf{H} = \mathbf{j}_{ext} - i\omega\left(\mathbf{P}_c + \mathbf{P}_{L,h} + \mathbf{P}_{NL,h}\right), \tag{4b}$$

where $\mathbf{P}_c + \mathbf{P}_{L,h} + \mathbf{P}_{NL,h} = \langle\varepsilon\mathbf{e}\rangle$, being $\mathbf{P}_c = \langle(\varepsilon-\varepsilon_h)\mathbf{e}\rangle$ the contribution to the macroscopic polarization due to the conduction currents in the nanowires, and $\mathbf{P}_{L,h} = \langle\varepsilon_h^0\mathbf{e}\rangle = \varepsilon_h^0\mathbf{E}$ and $\mathbf{P}_{NL,h} = \langle(\varepsilon_h - \varepsilon_h^0)\mathbf{e}\rangle = \varepsilon_h^0\langle\delta\varepsilon\mathbf{e}\rangle$ represent the contributions to the macroscopic polarization from the linear and nonlinear polarization currents in the host region, respectively. In the previous formulas, it is implicit that $\varepsilon - \varepsilon_h \equiv \varepsilon_m - \varepsilon_h^0$ in the metal region, and that $\delta\varepsilon$ is nonzero only in the dielectric region. We assume that the radius of the wires is small compared to period, and in these conditions it is typically possible to neglect the transverse part of the conduction polarization current, so that $\mathbf{P}_c = P_c\hat{\mathbf{z}}$. The conduction density of current in the metal may be written as $I/A_c = -i\omega P_c$, where $A_c = a^2$ is the area of the unit cell and $I$ can be identified with the microscopic current along the metallic wires (interpolated in a such a manner that it is defined over all the space) [25]. Therefore, Eq. (4b) may be rewritten as,

$$\nabla \times \mathbf{H} = \mathbf{j}_{ext} - i\omega\varepsilon_h^0 \mathbf{E} + \frac{I}{A_c}\hat{\mathbf{z}} - i\omega\mathbf{P}_{NL,h}. \tag{4b'}$$

Evidently, the conduction current satisfies a continuity equation

$$\frac{\partial I}{\partial z} = i\omega\sigma_l, \tag{5a}$$

where $\sigma_l$ is the charge density per unit of length (p.u.l.) on a nanowire [25]. Similar to what was done in Ref. [25], in order to describe the dynamics of $I$ from a macroscopic point of view, it is convenient to introduce a quasi-static potential $\varphi_w$. This additional potential may be regarded as the average potential drop from a given wire to the boundary of the respective unit cell [25] (see Appendix A), and is created by the electric charges on the metallic nanowires. It was shown in Ref. [25] that for the case of straight nanowires $\varphi_w$ satisfies:

$$\frac{\partial \varphi_w}{\partial z} = -(Z_w - i\omega L)I + E_z. \tag{5b}$$

This result remains valid in case of a nonlinear host. In the above, $L$ is a p.u.l. geometrical inductance given by

$$L = \frac{\mu_0}{2\pi}\log\left(\frac{a^2}{4r_w(a-r_w)}\right), \tag{6}$$

and $Z_w = -\dfrac{1}{i\omega\pi r_w^2(\varepsilon_m - \varepsilon_h^0)}$ is the p.u.l self-impedance of the nanowires. In the usual case wherein the metal permittivity follows a Drude dispersion model with plasma frequency

$\omega_m$ and collision frequency $\Gamma$, to a good approximation we can write

$$\varepsilon_m - \varepsilon_h^0 \sim \varepsilon_0 \frac{-\omega_m^2}{\omega(\omega+i\Gamma)}, \quad \text{and hence} \quad Z_w = R_{kin} - i\omega L_{kin} \quad \text{with} \quad R_{kin} = \frac{\Gamma}{\varepsilon_0 \omega_m^2 \pi r_w^2} \quad \text{and}$$

$L_{kin} = \frac{1}{\varepsilon_0 \omega_m^2 \pi r_w^2}$. The parameter $L_{kin}$ is the so-called kinetic inductance of the electrons in the metal. It is proven in Appendix A, that for the case of a (weakly) nonlinear host the linear density of charge $\sigma_l$ can be written in terms of the additional potential $\varphi_w$ as follows,

$$\sigma_l = C^0 \varphi_w + \frac{\varepsilon_h^0 A_c}{\varphi_w} \langle \delta\varepsilon \, \mathbf{e} \cdot \mathbf{e}_\Delta \rangle. \tag{7}$$

where $C^0$ denotes the p.u.l capacitance for straight nanowires in the linear problem [25]:

$$C^0 = \varepsilon_h^0 \left[ \frac{1}{2\pi} \log\left( \frac{a^2}{4r_w(a-r_w)} \right) \right]^{-1}. \tag{8}$$

The second term in Eq. (7) is evidently due to the nonlinear effects, and depends on the fluctuating part of the electric field, $\mathbf{e}_\Delta$ [see Eq. (3)].

Using Eq. (7), the system of Eqs. (5) can be rewritten as:

$$\frac{\partial I}{\partial z} = i\omega C^0 \varphi_w + i\omega \varepsilon_h^0 \frac{A_c}{\varphi_w} \langle \delta\varepsilon \, \mathbf{e} \cdot \mathbf{e}_\Delta \rangle, \tag{9a}$$

$$\frac{\partial \varphi_w}{\partial z} = i\omega L \zeta_w I + E_z. \tag{9b}$$

In the above, we defined the normalized impedance $\zeta_w = 1 + \frac{Z_w}{-i\omega L}$. For lossless wires ($R_{kin} = 0$) we have $\zeta_w = 1 + L_{kin}/L$, and for perfectly conducting wires $\zeta_w = 1$. Note that the normalized impedance $\zeta_w$ depends exclusively on the properties of the metal and on the geometry of the wire medium.

The system formed by the equations (4a)-(4b') and (9a)-(9b) describes the effective response of the nanowire array in terms of the macroscopic state-variables, $\mathbf{E}$, $\mathbf{H}$, $I$ and $\varphi_w$. Evidently, for a linear problem $\mathbf{P}_{NL,h} = 0$ and $\langle \delta\varepsilon \, \mathbf{e} \cdot \mathbf{e}_\Delta \rangle = 0$, and in such a case the system (4a)-(4b') and (9a)-(9b) is equivalent to the model of Ref. [25]. Here, the main challenge is to write $\mathbf{P}_{NL,h}$ and $\langle \delta\varepsilon \, \mathbf{e} \cdot \mathbf{e}_\Delta \rangle$ in terms of the macroscopic state-variables. This will be done in the next section.

It is interesting to mention that in case of a linear host, one can solve (9a)-(9b) with respect to $P_c = I/(-i\omega A_c)$ to find that:

$$P_c = -\varepsilon_h^0 \beta_p^2 \left( \frac{\partial^2}{\partial z^2} + k_h^2 \zeta_w \right)^{-1} E_z. \tag{10}$$

We defined $k_h^2 = \omega^2 \mu_0 \varepsilon_h^0$ and $\beta_p^2 = \mu_0/(LA_c) = C^0/(\varepsilon_h^0 A_c)$ so that:

$$\beta_p = \frac{1}{a} \left[ \frac{1}{2\pi} \log\left( \frac{a^2}{4r_w(a-r_w)} \right) \right]^{-1/2}. \tag{11}$$

The parameter $\beta_p$ is the "geometrical component" of the effective *plasma-wave number* of the metamaterial [23, 24, 25]. Thus, the polarization component ($P_c$) associated with the drift of electrons in the metal is related to the macroscopic electric field through an integral equation. This is the reason for the strongly spatially dispersive response [35] of arrays of metallic wires, which has been extensively discussed and highlighted in the literature [23, 24, 25]. Note that for plane wave propagation, one has the correspondence $\frac{\partial}{\partial z} \leftrightarrow ik_z$. As will be discussed ahead, in the nonlinear problem it is not possible to eliminate the additional variables $I$ and $\varphi_w$, and thus the macroscopic response of the system must be described by the eight-component $(\mathbf{E}, \mathbf{H}, I, \varphi_w)$ state vector, rather than by only $(\mathbf{E}, \mathbf{H})$ as in conventional media.

## III. Effective medium model

In order that Eqs. (4a)-(4b') and (9a)-(9b) form a closed effective medium model for the array of nanowires embedded in a nonlinear host, it is necessary to express the terms $\mathbf{P}_{NL,h} = \varepsilon_h^0 \langle \delta\varepsilon\, \mathbf{e} \rangle$ and $\langle \delta\varepsilon\, \mathbf{e} \cdot \mathbf{e}_\Delta \rangle$ as a function of the state variables, $(\mathbf{E}, \varphi_w)$. This is done in the following sub-sections.

### A. The nonlinear component of the polarization current in the host medium

To write $\mathbf{P}_{NL,h} = \varepsilon_h^0 \langle \delta\varepsilon\, \mathbf{e} \rangle$ in terms of the macroscopic fields, first we note that in the dielectric region $\delta\varepsilon\mathbf{e} = \alpha |\mathbf{e}|^2 \mathbf{e}$ is proportional to $\alpha$, and thus it is also proportional to the $\chi^{(3)}$ parameter. Let us suppose that somehow we are able to write $\langle |\mathbf{e}|^2 \mathbf{e} \rangle$ as a function of the macroscopic fields for the case of a linear problem (i.e. for $\alpha = 0$). Then, for a weak nonlinearity the following should be valid

$$\mathbf{P}_{NL,h} \approx \varepsilon_h^0 \alpha \langle |\mathbf{e}|^2 \mathbf{e} \rangle_L, \tag{12}$$

where the subscript $L$ in $\langle\ \rangle_L$ indicates that the spatial averaging of $|\mathbf{e}|^2 \mathbf{e}$ should be determined for the *linear problem*. Thus, within this approximation, all it is required is to determine $\langle |\mathbf{e}|^2 \mathbf{e} \rangle_L$ in terms of $(\mathbf{E}, \varphi_w)$ for a linear problem. It is implicit in the previous formula and in other equations of this section that the argument of the averaging operator ($|\mathbf{e}|^2 \mathbf{e}$) should be taken equal to zero in the metal region, because $\delta \varepsilon \mathbf{e} = \alpha |\mathbf{e}|^2 \mathbf{e}$ is non-zero only in the dielectric. In practice, it may not make much difference to consider that $\delta \varepsilon \mathbf{e} = \alpha |\mathbf{e}|^2 \mathbf{e}$ holds in the entire unit cell, because for a good conductor the electric field is very weak in the metal region.

To determine $\left\langle |\mathbf{e}|^2 \mathbf{e} \right\rangle_L$ in terms of $(\mathbf{E}, \varphi_w)$, we use the decomposition $\mathbf{e} = \mathbf{E} + \mathbf{e}_\Delta$ [Eq. (3)] to write:

$$\left\langle |\mathbf{e}|^2 \mathbf{e} \right\rangle_L = \left\langle |\mathbf{e}|^2 \right\rangle_L \mathbf{E} + \left\langle |\mathbf{e}|^2 \mathbf{e}_\Delta \right\rangle_L. \tag{13}$$

We used the fact that $\langle f \mathbf{F} \rangle = \langle f \rangle \mathbf{F}$, which is valid for a generic functions $f$ and $\mathbf{F}$ when $\mathbf{F}$ is a "macroscopic" field that varies slowly on the scale of the unit cell. It is proven in Appendix B that [Eqs. (B4) and (B10)]:

$$\left\langle \mathbf{e} \cdot \mathbf{e}^* \right\rangle_L = \mathbf{E} \cdot \mathbf{E}^* + \beta_p^2 |\varphi_w|^2, \tag{14}$$

$$\left\langle |\mathbf{e}|^2 \mathbf{e}_\Delta \right\rangle_L = \frac{1}{2} \beta_p^2 \varphi_w^2 \mathbf{E}_t^* + \frac{1}{2} \beta_p^2 |\varphi_w|^2 \mathbf{E}_t, \tag{15}$$

where $\mathbf{E}_t = E_x \hat{\mathbf{x}} + E_y \hat{\mathbf{y}}$ is the transverse (to $z$) part of the macroscopic electric field. Substituting these results into Eq. (12), we obtain the desired formula for the nonlinear component of the macroscopic polarization current:

$$\mathbf{P}_{NL,h} = \varepsilon_h^0 \alpha \left( \mathbf{E} \cdot \mathbf{E}^* + \beta_p^2 |\varphi_w|^2 \right) \mathbf{E} + \varepsilon_h^0 \frac{\alpha}{2} \beta_p^2 \left( \varphi_w^2 \mathbf{E}_t^* + |\varphi_w|^2 \mathbf{E}_t \right). \tag{16}$$

## B. The nonlinear component of the charge density

Next, we determine the nonlinear component of the charge density in terms of the macroscopic fields [see Eq. (7)]. To do this we note that $\delta\varepsilon\, \mathbf{e}\cdot\mathbf{e}_\Delta = \alpha |\mathbf{e}|^2 \mathbf{e}\cdot\mathbf{e}_\Delta$ in the host region, and hence using arguments analogous to those of Sect. III.A it is found that:

$$\langle \delta\varepsilon\, \mathbf{e}\cdot\mathbf{e}_\Delta \rangle \approx \alpha \langle |\mathbf{e}|^2 \mathbf{e}\cdot\mathbf{e}_\Delta \rangle_L. \tag{17}$$

Using repeatedly $\mathbf{e} = \mathbf{E} + \mathbf{e}_\Delta$ and $\langle f\,\mathbf{F}\rangle = \langle f \rangle \mathbf{F}$ (see the previous sub-section), it is found that:

$$\frac{1}{\alpha}\langle \delta\varepsilon\, \mathbf{e}\cdot\mathbf{e}_\Delta \rangle = \langle |\mathbf{e}|^2 \mathbf{e}_\Delta \rangle_L \cdot \mathbf{E} + \langle \mathbf{e}_\Delta \cdot \mathbf{e}_\Delta \rangle_L \mathbf{E}^* \cdot \mathbf{E} + \\ \langle |\mathbf{e}_\Delta|^2 \mathbf{e}_\Delta \cdot \mathbf{e}_\Delta \rangle_L + \langle \mathbf{e}_\Delta \cdot \mathbf{e}_\Delta \mathbf{e}_\Delta \rangle_L \cdot \mathbf{E}^* + \langle \mathbf{e}_\Delta \cdot \mathbf{e}_\Delta \mathbf{e}_\Delta^* \rangle_L \cdot \mathbf{E} \tag{18}$$

It is proven in Appendix B that $\langle \mathbf{e}_\Delta \cdot \mathbf{e}_\Delta \mathbf{e}_\Delta^* \rangle_L \approx 0 \approx \langle \mathbf{e}_\Delta \cdot \mathbf{e}_\Delta \mathbf{e}_\Delta \rangle_L$ [Eq. (B9)] and that $\langle \mathbf{e}_\Delta \cdot \mathbf{e}_\Delta \rangle_L = \beta_p^2 \varphi_w^2$ [Eq. (B6)]. On the other hand, $\langle |\mathbf{e}_\Delta|^2 \mathbf{e}_\Delta \cdot \mathbf{e}_\Delta \rangle$ is given by [Eq. (B12)]:

$$\langle |\mathbf{e}_\Delta|^2 \mathbf{e}_\Delta \cdot \mathbf{e}_\Delta \rangle_L = \tilde{B}\, \beta_p^4 \varphi_w^* \varphi_w^3. \tag{19}$$

where $\tilde{B}$ is a dimensionless parameter that only depends on $r_w/a$, and is given by Eq. (B14). Thus, substituting these results and Eq. (15) into Eq. (18), we obtain the desired formula:

$$\frac{1}{\alpha}\langle \delta\varepsilon\, \mathbf{e}\cdot\mathbf{e}_\Delta \rangle = \frac{1}{2}\beta_p^2 \left( \varphi_w^2 \mathbf{E}_t^* + |\varphi_w|^2 \mathbf{E}_t \right)\cdot \mathbf{E}_t + \beta_p^2 \varphi_w^2 \mathbf{E}^* \cdot \mathbf{E} + \tilde{B}\,\beta_p^4 \varphi_w^* \varphi_w^3. \tag{20}$$

## C. Summary

Substituting Eqs. (16) and (20) into Eqs. (4a)-(4b') and (9a)-(9b), and using $\beta_p^2 = C^0/(\varepsilon_h^0 A_c)$, it is found that the macroscopic electromagnetic fields in the nanowire metamaterial satisfy the following nonlinear partial differential system:

$$\nabla \times \mathbf{E} = +i\omega\mu_0 \mathbf{H}, \tag{21a}$$

$$\nabla \times \mathbf{H} = \mathbf{j}_{ext} - i\omega\varepsilon_{ef,h}\mathbf{E} + \frac{I}{A_c}\hat{\mathbf{z}} + \frac{1}{A_c}\mathbf{Y}\varphi_w, \tag{21b}$$

$$\frac{\partial I}{\partial z} = i\omega C\varphi_w - \mathbf{Y}\cdot\mathbf{E}_t, \tag{21c}$$

$$\frac{\partial \varphi_w}{\partial z} = i\omega L \zeta_w I + E_z. \tag{21d}$$

where $\zeta_w = 1 + \frac{Z_w}{-i\omega L}$, and the parameters $\varepsilon_{ef,h}$, $C$, and $\mathbf{Y}$ are nonlinear functions of the macroscopic state variables given by:

$$\varepsilon_{ef,h} = \varepsilon_h^0\left[1 + \alpha\left(\mathbf{E}\cdot\mathbf{E}^* + \beta_p^2 \varphi_w \varphi_w^*\right)\right], \tag{22a}$$

$$C = C^0\left[1 + \alpha\left(\mathbf{E}^*\cdot\mathbf{E} + \tilde{B}\beta_p^2 \varphi_w \varphi_w^*\right)\right], \tag{22b}$$

$$\mathbf{Y} = -i\omega C^0 \frac{\alpha}{2}\left(\varphi_w \mathbf{E}_t^* + \varphi_w^* \mathbf{E}_t\right). \tag{22c}$$

The parameter $\varepsilon_{ef,h}$ may be regarded as the effective nonlinear permittivity of the host medium, $C$ as the effective nonlinear p.u.l. wire capacitance, and $\mathbf{Y}$ is a parameter with unities of $\left[\Omega^{-1}\right]$ that determines the contribution of the macroscopic field to the induced charge density in the metal. The differential system (21) provides a full description of the dynamics of the macroscopic state variables $(\mathbf{E}, \mathbf{H}, I, \varphi_w)$. It is interesting to point out that with the exception of the parameter $\tilde{B}$ [Eq. (B14)], all the remaining parameters of the model are the ones introduced in Ref. [25] for the linear case.

## IV. Plane wave modes

As a first application of the developed effective medium model, next we study the impact of the nonlinear effects on the isofrequency contours of the natural modes of the nanowire material. In subsection IV.A, we determine the effective dielectric function of the nonlinear structured material, and discuss in a qualitative way how the nonlinear response tailors the shape of the isofrequency contours. In subsection IV.B, we obtain a formula for the Poynting vector in the nanowire material.

### A. Nonlocal dielectric function

It is possible to partially eliminate the variables $I$ and $\varphi_w$ and obtain in this manner the nonlocal dielectric function of the nanowire material. To this end, we solve (21d) for the current $I$ to obtain $I = \left( \frac{\partial \varphi_w}{\partial z} - E_z \right) / (i\omega L \zeta_w)$. Substituting this result into (21b) it follows that:

$$\nabla \times \mathbf{H} = \mathbf{j}_{ext} - i\omega \varepsilon_{ef,h} \mathbf{E} + \frac{1}{A_c} \frac{1}{i\omega L \zeta_w} \left( \frac{\partial \varphi_w}{\partial z} - E_z \right) \hat{\mathbf{z}} + \frac{1}{A_c} \mathbf{Y} \varphi_w. \tag{23}$$

Differentiating Eq. (21d) with respect to $z$ and using Eq. (21c), it is found that $\varphi_w$ satisfies the second order (nonlinear) differential equation:

$$\frac{\partial^2 \varphi_w}{\partial z^2} + \omega^2 L C \zeta_w \varphi_w = -i\omega L \zeta_w \mathbf{Y} \cdot \mathbf{E}_t + \frac{\partial E_z}{\partial z}. \tag{24}$$

We define $n_w^2$ as,

$$n_w^2 = \frac{C}{C^0} \zeta_w = \zeta_w \left[ 1 + \alpha \left( \mathbf{E}^* \cdot \mathbf{E} + \tilde{B} \beta_p^2 \varphi_w \varphi_w^* \right) \right]. \tag{25}$$

Since $LC^0 = \varepsilon_h^0 \mu_0$ it follows that putting $ik_z \leftrightarrow d/dz$ and $k_h^2 = \omega^2 \varepsilon_h^0 \mu_0$:

$$\varphi_w = \left( k_h^2 n_w^2 - k_z^2 \right)^{-1} \left( -i\omega L \zeta_w \mathbf{Y} \cdot \mathbf{E}_t + ik_z E_z \right). \tag{26}$$

It is important to mention that the right-hand side depends on $\varphi_w$, because $n_w^2$ and $\mathbf{Y}$ also do. Substituting the previous result into Eq. (23), it is seen that the equation can be rewritten as $\nabla \times \mathbf{H} = \mathbf{j}_{ext} - i\omega \bar{\bar{\varepsilon}}(\omega, k_z) \cdot \mathbf{E}$ where the nonlocal dielectric function $\bar{\bar{\varepsilon}}(\omega, k_z)$ is defined by:

$$\bar{\bar{\varepsilon}}(\omega, k_z) = \varepsilon_{ef,h} \bar{\bar{\mathbf{I}}} - \frac{1}{\zeta_w} \frac{\varepsilon_h^0 \beta_p^2}{\left(k_h^2 - k_z^2 / n_w^2\right)} \hat{\mathbf{z}}\hat{\mathbf{z}}$$
$$+ \frac{k_z}{\omega} \frac{1}{\left(k_h^2 n_w^2 - k_z^2\right)} \frac{1}{A_c} \left(\hat{\mathbf{z}}\mathbf{Y} - \mathbf{Y}\hat{\mathbf{z}}\right) + \frac{L\zeta_w}{A_c} \frac{1}{\left(k_h^2 n_w^2 - k_z^2\right)} \mathbf{Y}\mathbf{Y} \quad (27)$$

To obtain the above formula the identity $\beta_p^2 = \mu_0 / (LA_c)$ was used. The terms $\hat{\mathbf{z}}\hat{\mathbf{z}}, \hat{\mathbf{z}}\mathbf{Y}, \mathbf{Y}\hat{\mathbf{z}}, \mathbf{Y}\mathbf{Y}$ should be understood as tensor products of two vectors. The derived expression for the nonlocal dielectric function only applies to stationary solutions of Eq. (21) such that the dependence on $z$ of the fields is of the form $e^{ik_z z}$. In case of a linear host, the last two terms of the dielectric function vanish and $n_w^2 = \zeta_w$, so that the above result reduces to the formulas of Ref. [23, 24, 25]. It is interesting to note that despite the fact the host may be nonlinear, the nonlocal dielectric function satisfies $\bar{\bar{\varepsilon}}(\omega, k_z) = \bar{\bar{\varepsilon}}(\omega, -k_z)^T$, consistent with the usual properties of the nonlocal dielectric function in reciprocal media [35]. The superscript "$T$" refers to the transpose dyadic. Despite its elegance the formula (27) for the nonlocal dielectric function is of limited applicability, because due to the nonlinear effects the parameters $\varepsilon_{ef,h}$, $n_w^2$ and $\mathbf{Y}$ are all functions of $\varphi_w$ and $\mathbf{E}$. In particular, one sees that it is not possible to eliminate $\varphi_w$, so that $\bar{\bar{\varepsilon}}(\omega, k_z)$ depends exclusively on $\mathbf{E}$.

However, it is possible to considerably simplify Eq. (27) if one is only interested in electromagnetic field distributions such that $\varphi_w$ is $\pm 90°$ out of phase with respect to $\mathbf{E}_t$. This situation is of particular relevance because in the linear problem, all the plane wave natural modes associated with a real valued wave vector **k** have the enunciated property in case of negligible absorption loss. Indeed, one can see from Eq. (21c) that in the linear case and for a spatial variation of the form $e^{i\mathbf{k}\cdot\mathbf{r}}$, the current and the additional potential are in phase. Thus, from Eq. (21d) it is evident that $\varphi_w$ and $E_z$ are 90º out of phase in a lossless medium ($\zeta_w$ real valued). It is known that for propagating plane waves in the nanowire material all the components of the electric field are in phase [23, 24, 25], and hence the enunciated result follows. We will show ahead that in the nonlinear case, the medium also supports plane wave solutions such that $\varphi_w$ and $\mathbf{E}_t$ are 90º out of phase. For field distributions with this property it turns out that the admittance **Y** vanishes, and hence Eq. (27) reduces to the simple expression:

$$\frac{1}{\varepsilon_h^0}\overline{\overline{\varepsilon}}(\omega,k_z) = n_{ef,h}^2 \overline{\overline{\mathbf{I}}} - \frac{1}{\zeta_w}\frac{\beta_p^2}{\left(k_h^2 - k_z^2/n_w^2\right)}\hat{\mathbf{z}}\hat{\mathbf{z}}. \tag{28}$$

We defined $n_{ef,h}^2 = \varepsilon_{ef,h}/\varepsilon_h^0$ given by:

$$n_{ef,h}^2 = 1 + \alpha\left(\mathbf{E}\cdot\mathbf{E}^* + \beta_p^2 \varphi_w \varphi_w^*\right), \tag{29}$$

It is underlined that both $n_{ef,h}^2$ and $n_w^2$ are nonlinear functions of the macroscopic fields. One can qualitatively understand from Eq. (28) how the nonlinear response affects the wave propagation. Indeed, for the case of a self-focusing host $(\alpha > 0)$ the nonlinear effects necessarily result in an increase of $n_w^2$. This parameter can be identified with the slow wave factor of Ref. [25]. It was shown in that work that an increase of $n_w^2$ reduces the effects of spatial dispersion. In short, the reason is that the dependence on $k_z$ of the second term in the right-hand side of Eq. (28) is of the form $k_z^2/n_w^2$ and thus it is less significant for larger values of the slow wave factor. Based on this discussion, one can expect that if the host of the nanowires is a self-focusing Kerr medium the effects of spatial dispersion may be slightly tamed for high field intensities, and the metamaterial may behave closer to a local hyperbolic (indefinite) medium.

To illustrate this, next we study the isofrequency contours for the plane wave modes with transverse magnetic (TM)-polarization (extraordinary wave in the uniaxial medium). The isofrequency contours are the solutions of the characteristic equation $k_t^2 \frac{1}{\varepsilon_{zz}} + k_z^2 \frac{1}{\varepsilon_t} = \left(\frac{\omega}{c}\right)^2$ with $\varepsilon_t \equiv \varepsilon_{xx} = \varepsilon_{yy}$ and $k_t^2 = k_x^2 + k_y^2$. From Eq. (28), after some simplifications, the characteristic equation can be written as:

$$k_t^2 + k_z^2 - n_{ef,h}^2 k_h^2 + \frac{\beta_p^2}{\zeta_w} \frac{k_h^2 - k_z^2/n_{ef,h}^2}{k_h^2 - k_z^2/n_w^2} = 0. \tag{30}$$

As discussed previously, $n_{ef,h}^2$ and $n_w^2$ depend on the macroscopic fields, but in the following discussion they will be treated as constant parameters. The characteristic equation can be reduced to a polynomial equation quadratic in $k_z^2$, and hence there are two relevant electromagnetic modes, usually designated by quasi-transverse electromagnetic (q-TEM) and TM modes [24, 25]. Here, we are only interested in the isofrequency contours of the q-TEM mode, which has no lower frequency cut-off and is the only propagating mode for frequencies $f < f_{p,ef}$, being $f_{p,ef}$ the effective plasma frequency defined so that $\varepsilon_{zz}(2\pi f_{p,ef}, 0) = 0$ in the linear case. It satisfies $2\pi f_{p,ef} \sqrt{\mu_0 \varepsilon_h^0} = \beta_p / \sqrt{\zeta_w}$. Figure 2 shows the computed isofrequency contours for different nanowire arrays considering both the linear case (solid lines, obtained with $n_{ef,h}^2 = 1.0$ and $n_w^2 = \zeta_w$) and the nonlinear case (dashed lines) for a self-focusing host medium. The nonlinear case is modeled by assuming $n_{ef,h}^2 = 1.01$ and $n_w^2 = \zeta_w (1+0.2)$. We take $n_w^2 / \zeta_w > n_{ef,h}^2$ [see Eqs. (25) and (29)] because the parameter $\tilde{B}$ is typically significantly larger than unity (e.g. for $r_w/a = 0.05$ we have $\tilde{B} = 7.7$). The value $n_w^2 / \zeta_w$ considered here is probably too high to be achieved in practice, but here we just want to discuss qualitatively the effect of the Kerr-type nonlinearity, and the difference between the linear and the nonlinear cases is clearer with a larger value of $n_w^2 / \zeta_w$. It is assumed that the nanowires are made of silver, which is modeled by $\omega_m / 2\pi = 2175 [\text{THz}]$ [36]. For simplicity, we neglect the effect of metal loss, which is typically negligible provided the radius of the nanowires is much larger than the metal skin depth ($\delta_m$). This condition is met to a reasonable approximation in common designs in the terahertz regime and at lower frequencies.

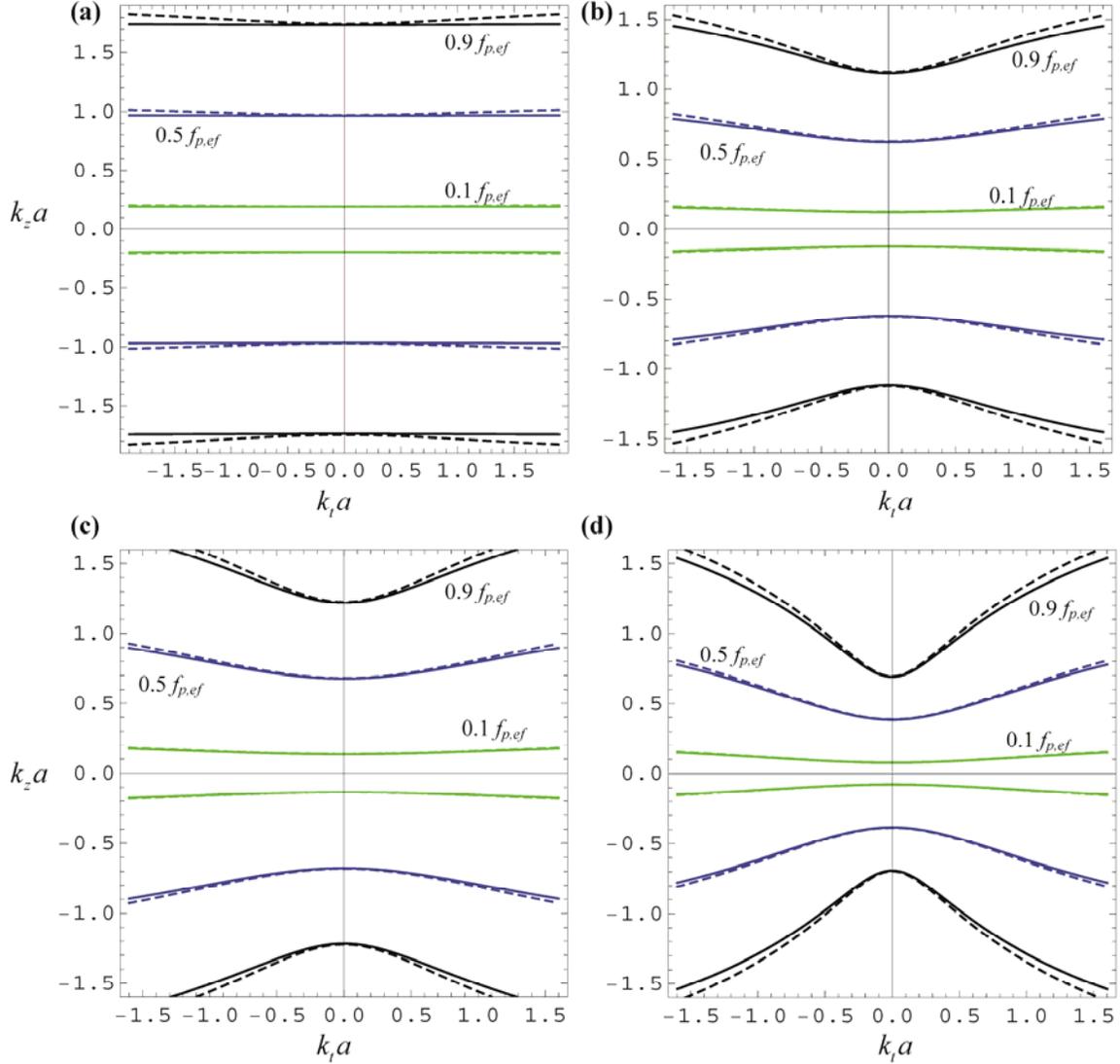

Fig. 2. (Color online) Isofrequency contours for the extraordinary (q-TEM) wave in a nanowire material formed by silver wires in a dielectric background with $\varepsilon_h^0 = 1.0\varepsilon_0$. (a) $r_w = 200nm$, $r_w/a = 0.05$; (b) $r_w = 20nm$, $r_w/a = 0.05$; (c) $r_w = 20nm$, $r_w/a = 0.1$; (d) $r_w = 10nm$, $r_w/a = 0.1$. The insets in the plots indicate the value of the frequency of operation. The effective plasma frequency ($f_{p,ef}$) is equal to 23THz, 148 THz, 323 THz, and 367 THz for the plots (a)-(d), respectively. The solid lines are computed for $n_{ef,h}^2 = 1.0$ and $n_w^2 = \zeta_w$ and the dashed lines for $n_{ef,h}^2 = 1.01$ and $n_w^2 = \zeta_w(1+0.2)$.

The results of Fig. 2 demonstrate that in all the examples the nonlinear effects enhance the hyperbolic shape of the isofrequency contours. This is especially evident for frequencies comparable with $f_{p,ef}$ (black lines), and for the case of thin wires and dense arrays (example (d)). Notice that the isofrequency contours become nearly flat when the radius of the wires is much larger than the skin depth of the metal (example (a)), or equivalently if $L_{kin} \ll L$. In such a case, the effective medium is characterized by a very strong anisotropy and the negative refraction effect is quite weak. When the nanowires radius becomes comparable or smaller than $\delta_m$ a strong indefinite response is obtained. This property has been discussed in previous works [37, 38].

## B. The Poynting vector

In Ref. [39] it was shown that the Poynting vector in a nanowire material with a linear dielectric host can be expressed in terms of the macroscopic variables as follows (the result below assumes a time harmonic excitation):

$$\mathbf{S} = \frac{1}{2}\mathrm{Re}\left\{ \mathbf{E} \times \mathbf{H}^* + \frac{1}{A_c}\varphi_w I^* \hat{\mathbf{z}} \right\}. \tag{31}$$

In this subsection, we argue that this formula for the time averaged Poynting vector also holds in the nonlinear case. For simplicity in what follows the nanowire material is assumed lossless. This implies that $\varepsilon_{ef,h}$, $C$ and $\zeta_w$ are real-valued. Straightforward calculations based on Eqs. (21) show that:

$$\nabla \cdot \mathbf{S} = \frac{1}{2}\mathrm{Re}\left\{ -\mathbf{E}^* \cdot \mathbf{j}_{ext} - \frac{1}{A_c}\mathbf{E}^* \cdot \mathbf{Y}\varphi_w - \frac{1}{A_c}\mathbf{Y} \cdot \mathbf{E}_t \varphi_w^* \right\}. \tag{32}$$

From Eq. (22c) it should be clear that for a lossless system $\mathbf{Y}$ is imaginary pure. Hence, $\mathbf{E}^* \cdot \mathbf{Y}\varphi_w + \mathbf{Y} \cdot \mathbf{E}_t \varphi_w^* = -\left(\mathbf{Y} \cdot \mathbf{E}_t\right)^* \varphi_w + \mathbf{Y} \cdot \mathbf{E}_t \varphi_w^* = 2i\,\mathrm{Im}\left\{\mathbf{Y} \cdot \mathbf{E}_t \varphi_w^*\right\}$. Thus, it follows that:

$$\nabla \cdot \mathbf{S} = \frac{1}{2}\mathrm{Re}\left\{ -\mathbf{E}^* \cdot \mathbf{j}_{ext} \right\}. \tag{33}$$

The right-hand side of the above equation corresponds to the average power density extracted from the external sources, and hence **S** defined by Eq. (31) is indeed consistent with the formula for the macroscopic Poynting vector in the steady-state regime. Obviously, in case of a lossy system the right-hand side of Eq. (33) must be modified to include the terms related to the absorption loss in the materials.

The Poynting vector can be written in terms of $(\mathbf{E}, \varphi_w)$. To this end, we use (21a) and (21d) to eliminate $(\mathbf{H}, I)$ and obtain:

$$\mathbf{S} = \frac{1}{2\omega\mu_0} \operatorname{Re}\left\{ \mathbf{E} \times \left(\frac{1}{i}\nabla \times \mathbf{E}\right)^* + \frac{\beta_p^2}{\zeta_w^*} \varphi_w \left(\frac{1}{i}\frac{\partial \varphi_w}{\partial z}\right)^* \hat{\mathbf{z}} - \frac{\beta_p^2}{\zeta_w} i\varphi_w E_z^* \hat{\mathbf{z}} \right\}. \qquad (34)$$

In case the electromagnetic fields have a spatial dependence of the form $e^{i\mathbf{k}\cdot\mathbf{r}}$ the Poynting vector becomes:

$$\mathbf{S} = \frac{1}{2k_0\eta_0} \operatorname{Re}\left\{ \mathbf{E} \times (\mathbf{k} \times \mathbf{E})^* + \frac{\beta_p^2}{\zeta_w^*} k_z^* |\varphi_w|^2 \hat{\mathbf{z}} - \frac{\beta_p^2}{\zeta_w} i\varphi_w E_z^* \hat{\mathbf{z}} \right\}, \qquad (35)$$

where $k_0 = \omega/c = \omega\sqrt{\mu_0\varepsilon_0}$ is the free-space wave number and $\eta_0 = \sqrt{\mu_0/\varepsilon_0}$ is the intrinsic impedance of vacuum.

## V. Refraction by the Nanowire Material

Next, we study the refraction of an incoming plane wave propagating in free-space by a semi-infinite nanowire array embedded in a nonlinear Kerr-type medium (Fig. 3). The objective is to characterize the angle of transmission $\theta_t$ at an air − nanowire array interface as a function of the intensity of the transmitted Poynting vector. It is assumed that the plane of incidence is the *yoz* plane and that the incoming electric field has TM-polarization (the only non-zero components of the electric field are $E_y$ and $E_z$, and the fields are independent of the $x$ coordinate so that $\partial_x = 0$). The incident wave varies with $y$ (direction parallel to the interface) as $e^{ik_y y}$, being $k_y = (\omega/c)\sin\theta_i$ determined by the angle of incidence $\theta_i$. Thus, it should be clear that within the continuous medium approximation [Eq. (21)], the state variables $(\mathbf{E}, \mathbf{H}, \varphi_w, I)$ also depend on $y$ as $e^{ik_y y}$ in all space.

The exact solution of Eq. (21) in the described scenario of refraction can only be obtained by a numerical approach, because in principle, due to the nonlinear effects, $\varepsilon_{ef,h}$, $n_w^2$ and $\mathbf{Y}$ can be complicated (continuous) functions of $z$ near the interface. However, sufficiently far away from the interface it seems reasonable to suppose that if the loss in the material is negligible, the transmitted field may reduce to a propagating plane wave mode with constant amplitude. In particular, sufficiently deep inside the nanowire material (after some transition layer) the parameters $\varepsilon_{ef,h}$, $n_w^2$ and $\mathbf{Y}$ are expected become constants, whose value depends on the amplitude of the incident field. The analysis of this section is based on this hypothesis. In the following we determine the angle of transmission $\theta_t$ (Fig. 3; the angle is calculated after the mentioned transition layer) as a function of the energy density flux in the nanowire material. The angle $\theta_t$ is determined by the direction of the Poynting vector $\mathbf{S}$ in the nanowire array, and the energy density flux is $|\mathbf{S}|$. Notice that the actual value of $\mathbf{S}$ depends on the Poynting vector $\mathbf{S}^{inc}$ of the incoming wave. The exact relation between the amplitude of the Poynting vector of the transmitted wave ($\mathbf{S}^{tx}$) and that of the incident wave ($\mathbf{S}^{inc}$), determined by a reflectivity $R$, is out of the scope of the present work. To a first

-22-

approximation $R$ can be computed assuming a linear response, but this will not be discussed here.

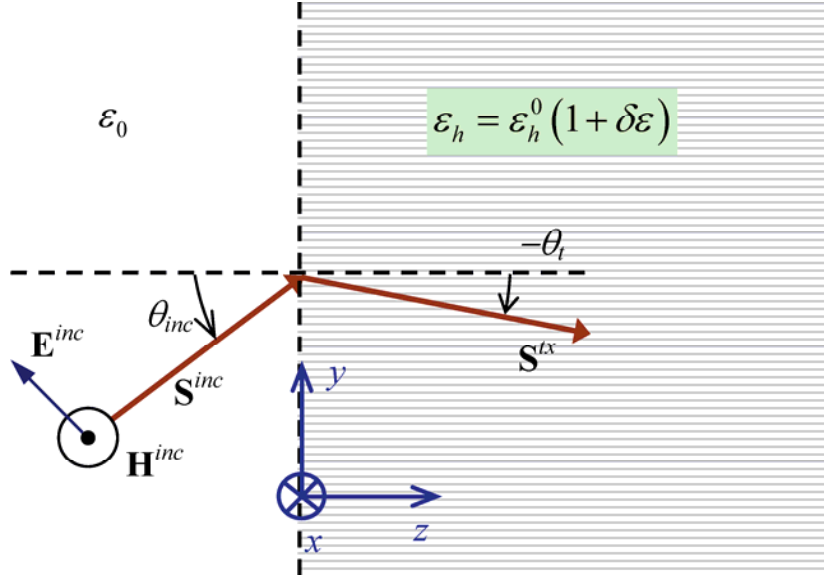

Fig. 3. (Color online) Refraction of a TM-polarized plane wave propagating in free-space by a semi-infinite nanowire material with a Kerr-type (self-focusing) nonlinear host. The nanowires are oriented along the $z$-direction. Sufficiently deep inside the nanowire material and in case of negligible absorption loss, the transmitted wave is expected to be a propagating plane wave characterized by the Poynting vector $\mathbf{S}^{tx}$. As illustrated in the figure, due to the hyperbolic nature of the isofrequency contours, the incoming wave is negatively refracted at the interface.

### *A. Formulation*

Next, we explain how the relation $\theta_t$ vs. $S^{tx}$ can be found. To begin with, the first order differential system (21) is reduced to a second order nonlinear differential system in $(\mathbf{E}, \varphi_w)$. To do this we use Eqs. (21a), (23) and (24) and $\beta_p^2 = \mu_0/(LA_c)$, to obtain:

$$\nabla \times \nabla \times \mathbf{E} - k_h^2 n_{ef,h}^2 \mathbf{E} = i\omega\mu_0 \mathbf{j}_{ext} + \frac{\beta_p^2}{\zeta_w}\left(\frac{\partial \varphi_w}{\partial z} - E_z\right)\hat{\mathbf{z}} + \beta_p^2 k_h^2 \tilde{\mathbf{Y}}\varphi_w, \qquad (36a)$$

$$\frac{\partial^2 \varphi_w}{\partial z^2} + k_h^2 n_w^2 \varphi_w = -\zeta_w k_h^2 \tilde{\mathbf{Y}} \cdot \mathbf{E}_t + \frac{\partial E_z}{\partial z}. \qquad (36b)$$

We defined $k_h^2 \tilde{\mathbf{Y}} = i\omega L \mathbf{Y}$ so that, taking into account that $LC^0 = \varepsilon_h^0 \mu_0$, we have $\tilde{\mathbf{Y}} = \frac{\alpha}{2}(\varphi_w \mathbf{E}_t^* + \varphi_w^* \mathbf{E}_t)$. In the same manner as in Sect. IV.A, we look for natural modes ($\mathbf{j}_{ext} = 0$) such that $\varphi_w$ is $\pm 90°$ out of phase with respect to $\mathbf{E}_t$ so that $\tilde{\mathbf{Y}} = 0$. Moreover, based on the geometry of the scattering problem (Fig. 3), we suppose that $\mathbf{E} = (0, E_y, E_z)$ and $\nabla = i\mathbf{k} = i(0, k_y, k_z)$. In these conditions, straightforward simplifications show that Eqs. (36) can be rewritten in a matrix form:

$$\begin{pmatrix} k_z^2 - k_h^2 n_{ef,h}^2 & -k_y k_z & 0 \\ -k_z k_y & k_y^2 - k_h^2 n_{ef,h}^2 + \dfrac{\beta_p^2}{\zeta_w} & -\dfrac{\beta_p^2}{\zeta_w} k_z a \\ 0 & -k_z & \left(k_z^2 - k_h^2 n_w^2\right) a \end{pmatrix} \begin{pmatrix} E_y \\ E_z \\ i\varphi_w / a \end{pmatrix} = 0. \tag{37}$$

Evidently, the above system is still nonlinear because $n_w^2$ and $n_{ef,h}^2$ are functions of $(\mathbf{E}, \varphi_w)$. Since by hypothesis $\varphi_w$ is $\pm 90°$ out of phase with respect to $\mathbf{E}_t$, the vector $\begin{pmatrix} E_y & E_z & i\varphi_w/a \end{pmatrix}^T$ can always be assumed real-valued. Thus, for convenience we write:

$$\begin{pmatrix} E_y \\ E_z \\ i\varphi_w / a \end{pmatrix} = A \begin{pmatrix} \cos\varphi \sin\theta \\ \sin\varphi \sin\theta \\ \cos\theta \end{pmatrix} \tag{38}$$

where $A$, $\theta$, and $\varphi$ are unknown and real-valued. The parameters $n_w^2$ and $n_{ef,h}^2$ [see Eqs. (25) and (29)] can be written in terms of $\theta$ and $\tilde{\alpha} = \alpha A^2$ as follows:

$$n_{ef,h}^2 = 1 + \tilde{\alpha}\left(\sin^2\theta + \beta_p^2 a^2 \cos^2\theta\right), \tag{39a}$$

$$n_w^2 = \zeta_w\left[1 + \tilde{\alpha}\left(\sin^2\theta + \tilde{B}\beta_p^2 a^2 \cos^2\theta\right)\right]. \tag{39b}$$

Substituting Eqs. (38)-(39) into the system (37), we obtain a nonlinear system of three equations of the form $\mathbf{F}(\theta,\varphi,k_z,\tilde{\alpha})=0$. The parameters $a,\beta_p,\zeta_w,\tilde{B}$ only depend on the geometry of the nanowire array and can be regarded as constants. On the other hand, $k_h$ is determined by the frequency of operation and $k_y=(\omega/c)\sin\theta_i$ by the angle of incidence.

To determine the relation $\theta_t$ vs. $S^{tx}$ we do the following. We arbitrarily fix the value of the small parameter $\tilde{\alpha}=\alpha A^2$ (which determines the strength of the weak nonlinear effects), and solve the nonlinear system of three equations $\mathbf{F}(\theta,\varphi,k_z,\tilde{\alpha})=0$ with respect to $(\theta,\varphi,k_z)$. After $(\theta,\varphi,k_z)$ is known, we use Eqs. (35) and (38) to calculate the Poynting vector $\mathbf{S}^{tx}$. Finally, we compute $S^{tx}=|\mathbf{S}^{tx}|$ and $\theta_t=\arctan^{-1}(S_y^{tx}/S_z^{tx})$. Repeating this procedure for large number of values of $\tilde{\alpha}$ one obtains a table with the values of $(\theta_t,S^{tx})$, which can be used to make the plots reported in the next sub-section.

It should be noted that because $\mathbf{F}(\theta,\varphi,k_z,\tilde{\alpha})=0$ is a nonlinear system in principle multiple solutions may exist for a fixed $\tilde{\alpha}$. For simplicity, in this work we only consider the solution which is obtained from the perturbation of the linear case ($\tilde{\alpha}=0$). Thus, in practice we start by solving $\mathbf{F}(\theta,\varphi,k_z,\tilde{\alpha})=0$ for the case $\tilde{\alpha}=0$. For $\tilde{\alpha}=0$ the problem can be reduced to a standard eigensystem. We use this solution (obtained for $\tilde{\alpha}=0$) as the initial guess in the root search of the nonlinear case.

### *B. Impact of the nonlinear response on the negative refraction*

Using the algorithm delineated in the previous subsection, we have determined the characteristic $\theta_t$ vs. $S^{tx}$ for different representative nanowire materials (Fig. 4). It is simple to check that $\theta_t$ only depends on the value of $\alpha S^{tx}$, being $\alpha$ the parameter that determines $\chi^{(3)}$. Hence, the horizontal axis of all the plots in Fig. 4 is taken equal to $2\eta_0 \alpha S^{tx}$ so that the Poynting vector is normalized to $\alpha$. Note that $2\eta_0 \alpha S^{tx}$ is roughly proportional to $\alpha |E^{tx}|^2 \sim \langle \delta\varepsilon \rangle$, i.e. it is of the order of the relative (spatially averaged) variation of the permittivity of the Kerr host dielectric. In the numerical calculations, similar to Sect. IV, it was assumed that the metallic wires are made of silver. The nanowires are embedded in a Kerr medium with $\varepsilon_h^0 = \varepsilon_0$ (a larger, and thus more realistic, value of $\varepsilon_h^0$ does not change appreciably the numerical results). The effects of absorption loss are not considered in our calculations. The typical values for the $\chi^{(3)}$ parameter of different crystals, glasses, polymers and liquids can be found in Ref. [32, p. 212]. For example, for Si $\chi^{(3)} = 2.8 \times 10^{-18} m^2/V^2$ and $\varepsilon_{h,r}^0 = 3.4^2$ so that $\alpha = 3\chi^{(3)}/\varepsilon_{h,r}^0 = 7.2 \times 10^{-19} m^2/V^2$ [32, 40]. Hence to have $2\eta_0 \alpha S^{tx} \sim 0.05$ one may estimate that the corresponding electric field amplitude is of the order of $2.6 \times 10^8 V/m$ which is achievable with a high intensity laser.

The plots of Fig. 4 predict an enhancement of the strength of negative refraction for high field intensities, particular for larger values of the incident angle $\theta_i$ (blue dot-dashed lines for $\theta_i = 40°$). This is consistent with the results of Fig. 2, because the nonlinear effects lead to a reduction of spatial dispersion and to isofrequency contours with greater hyperbolicity. The sensitivity to the nonlinearities is particularly strong for a frequency of operation closer to the effective plasma frequency (panels *iii* for $f = 0.9 f_{p,ef}$) and for the thinnest and denser nanowire arrays (panels *c*).

-26-

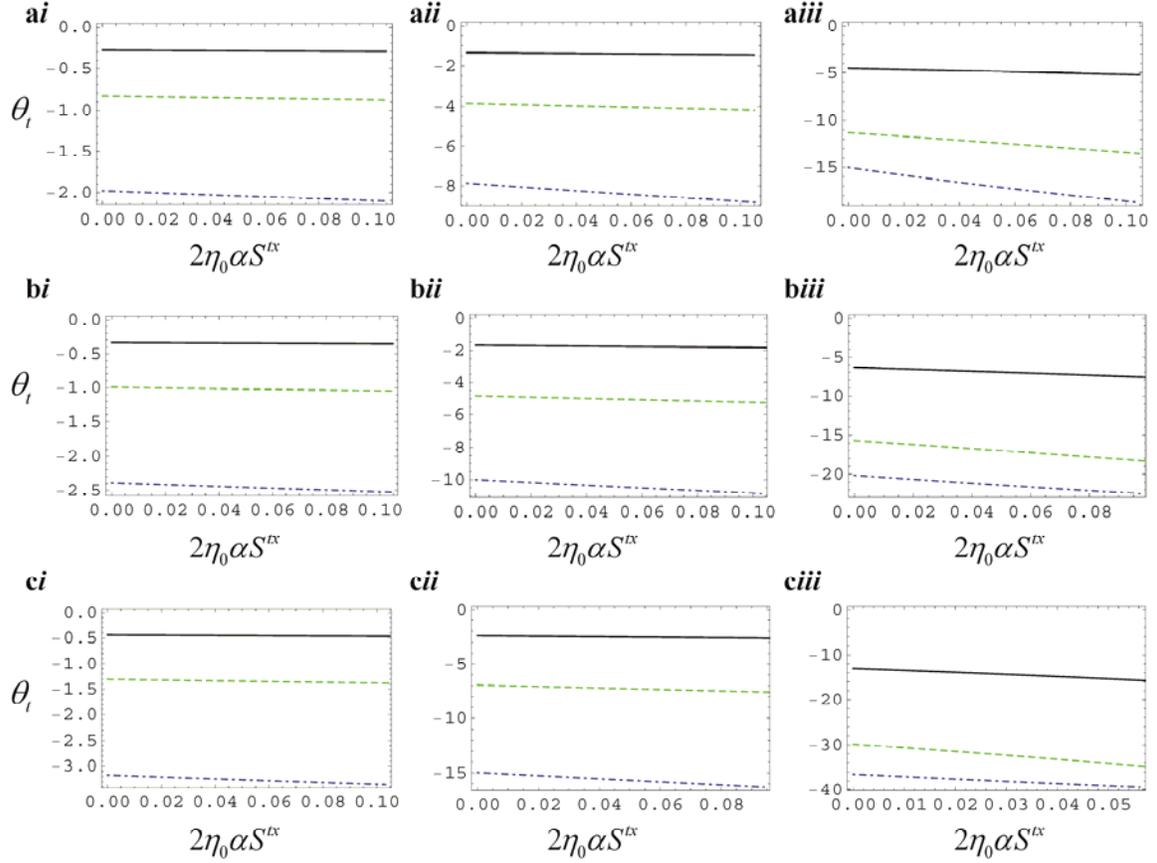

Fig. 4. (Color online) Transmission angle $\theta_t$ as a function of the normalized transmitted Poynting vector intensity ($S^{tx}$) in a nanowire material formed by silver wires embedded in a Kerr-type dielectric host with $\varepsilon_h^0 = 1.0\varepsilon_0$. (a) $r_w = 20nm$, $r_w/a = 0.05$; (b) $r_w = 20nm$, $r_w/a = 0.1$; (c) $r_w = 10nm$, $r_w/a = 0.1$. The cases (i), (ii), and (iii) are associated with the frequencies of operation $0.3 f_{p,ef}$, $0.6 f_{p,ef}$ and $0.9 f_{p,ef}$, being the effective plasma frequency equal to 148 THz, 323 THz, and 367 THz for the plots (a)-(c), respectively. Solid black lines: $\theta_i = 5°$; Green dashed lines: $\theta_i = 15°$; Blue dot-dashed lines: $\theta_i = 40°$.

# VI. Conclusion

We extended the effective medium model of a nanowire array developed in previous works [23, 24, 25] to the case wherein the host dielectric has a nonlinear Kerr-type response. The main result of this work is summarized by Eqs. (21)-(22). The macroscopic response of the structured material is described by the macroscopic state variables $(\mathbf{E}, \mathbf{H}, I, \varphi_w)$ that are coupled through a nonlinear differential system [Eq. (21)]. With the exception of the parameter $\tilde{B}$, all the parameters of the model were already known in the framework of the model developed in Ref. [25]. The proposed effective medium model applies in case of a weak nonlinear response [Eq. (12)], and is expected to be useful in the study of the modulation of the linear properties of the main harmonic due to nonlinear perturbations. Nonlinear processes involving very strong fields or two or more harmonics cannot be studied with the present theory. It should be emphasized that our derivations are based on the assumption of relatively thin wires, and that the electric field is typically stronger around the wire surface comparatively with the rest of the unit cell. Thus, the nonlinear response should be weak enough so that the nonlinear perturbations of the fields around the wire surface are tiny.

To illustrate the application of the theory, we characterized the low-frequency natural plane modes in the nanowire array, showing that the nonlinear effects lead to a reduction of spatial dispersion effects, and hyperbolic-like isofrequency contours. Our theory predicts an enhancement of the negative refraction at an air – nanowire array interface, when the nanowires are embedded in a self-focusing Kerr-type medium. The developed effective medium theory can be instrumental in the characterization of spatial solitons and other nonlinear effects.

**Acknowledgments:**

Useful discussions with S. I. Maslovski are gratefully acknowledged. This work is supported in part by Fundação para Ciência e a Tecnologia under project PTDC/EEATEL/100245/2008.

# Appendix A: The fluctuating part of the electric field and the additional potential

In this Appendix, we obtain explicit formulas for the fluctuating part of the microscopic electric field ($\mathbf{e}_\Delta$), for the additional potential ($\varphi_w$), and derive the relation between $\sigma_l$ and $\varphi_w$ [Eq. (7)] in case of a nonlinear host.

### I. *The fluctuating part of the microscopic electric field*

We assume that the microscopic electric field is a quasi-static field, and hence to a first approximation it can written in terms of a scalar potential $\phi = \phi(x, y, z)$:

$$\mathbf{e} = -\nabla \phi \tag{A1}$$

Evidently, the potential $\phi$ satisfies the equation $\nabla \cdot (\varepsilon \nabla \phi) = -\rho_{ext}$, which can be written as:

$$\nabla^2 \phi = -\frac{\rho_{ext}}{\varepsilon_h^0} - \frac{\rho_{NL,h}}{\varepsilon_h^0} - \frac{\rho_c}{\varepsilon_h^0}. \tag{A2}$$

where $\rho_{ext}$ is an external density of charge (due to an external macroscopic source, which can be trivial), $\rho_{NL,h} = \varepsilon_h^0 \nabla \cdot (\delta\varepsilon \nabla \phi)$ is the additional density of charge in the dielectric due to the nonlinear effects, and $\rho_c = \nabla \cdot \left((\varepsilon_m - \varepsilon_h^0) \nabla \phi\right)$ is the density of charge in the

-29-

metal (relative to the dielectric host). It is evident that the electrical potential has the following integral representation

$$\phi(\mathbf{r}) = \frac{1}{4\pi\varepsilon_h^0} \int \left[\rho_{ext}(\mathbf{r}') + \rho_{NL,h}(\mathbf{r}') + \rho_c(\mathbf{r}')\right] \frac{1}{|\mathbf{r}-\mathbf{r}'|} d^3\mathbf{r}', \qquad (A3)$$

where the integration is over all space.

Let us consider first that the field distribution corresponds to a Bloch wave associated with the wave vector $\mathbf{k} = (k_x, k_y, k_z)$. Note that the field is not necessarily a natural mode, because in general $\rho_{ext} \neq 0$. In such a case, using the Bloch wave property, it is possible to reduce the region of integration to the unit cell ($\Omega$) of the periodic structure. Indeed, we have:

$$\phi(\mathbf{r}) = \frac{1}{\varepsilon_h^0} \int_\Omega \left[\rho_{ext}(\mathbf{r}') + \rho_{NL,h}(\mathbf{r}') + \rho_c(\mathbf{r}')\right] \Phi(\mathbf{r}-\mathbf{r}') d^3\mathbf{r}', \qquad (A4)$$

where $\Phi(\mathbf{r}) = \sum_{\mathbf{I},n} e^{i\mathbf{k}\cdot\mathbf{r}_{\mathbf{I},n}} \frac{1}{4\pi|\mathbf{r}-\mathbf{r}_{\mathbf{I},n}|}$, and $\mathbf{r}_{\mathbf{I},n} = (a i_1, a i_2, a_z n)$ is a generic lattice point, $\mathbf{I} = (i_1, i_2)$ is a generic pair of integers, and $n$ is a generic integer. The period of the crystal along $z$ ($a_z$) can be chosen arbitrarily because the nanowire array is invariant to translations along $z$. The function $\Phi(\mathbf{r})$ satisfies $\nabla^2 \Phi = -\sum_{\mathbf{I},n} e^{i\mathbf{k}\cdot\mathbf{r}_{\mathbf{I},n}} \delta(\mathbf{r}-\mathbf{r}_{\mathbf{I},n})$, and thus can be regarded as a Green function. It has the following plane wave representation [41]:

$$\Phi(\mathbf{r}) = \frac{1}{A_c a_z} \sum_{\mathbf{J},n} \frac{1}{\mathbf{k}_\mathbf{J} \cdot \mathbf{k}_\mathbf{J} + \left(k_z + \frac{2\pi}{a_z} n\right)^2} e^{i\mathbf{k}_\mathbf{J}\cdot\mathbf{r}} e^{i\left(k_z + \frac{2\pi}{a_z}n\right)z}. \qquad (A5)$$

where $A_c = a^2$, $\mathbf{k}_\mathbf{J} = \mathbf{k}_t + \mathbf{k}_\mathbf{J}^0$, $\mathbf{k}_t = (k_x, k_y, 0)$, $\mathbf{k}_\mathbf{J}^0 = \left(j_1 \frac{2\pi}{a}, j_2 \frac{2\pi}{a}, 0\right)$ and $\mathbf{J} = (j_1, j_2)$ is a generic pair of integers. Since the nanowire crystal is invariant to translations along $z$ it is clear that for a Bloch wave $\rho_{ext}$, $\rho_{NL,h}$ and $\rho_c$ depend on $z$ as $e^{ik_z z}$. Thus, substituting the above formula into Eq. (A4), it is easily found that

$$\phi(\mathbf{r}) = \frac{1}{\varepsilon_h^0} e^{ik_z z} \int_{\Omega_t} \left[ \rho_{ext}(\mathbf{r}') + \rho_{NL,h}(\mathbf{r}') + \rho_c(\mathbf{r}') \right] e^{-ik_z z'} \Phi_{2D}(\mathbf{r} - \mathbf{r}') dx' dy', \tag{A6}$$

where $\Omega_t$ is the intersection of the unit cell with the *xoy* plane, and we introduced $\Phi_{2D}$ given by:

$$\Phi_{2D}(x, y) = \frac{1}{A_c} \sum_{\mathbf{J}} \frac{1}{\mathbf{k_J} \cdot \mathbf{k_J} + k_z^2} e^{i \mathbf{k_J} \cdot \mathbf{r}}. \tag{A7}$$

Next, we decompose the electrostatic potential into its spatial average and a fluctuating part: $\phi = \langle \phi \rangle + \phi_\Delta$. From the definition of the averaging operator (see Sect. II), it is easily seen that:

$$\langle \phi \rangle(\mathbf{r}) = \frac{1}{\varepsilon_h^0} \frac{1}{A_c} \frac{1}{\mathbf{k} \cdot \mathbf{k}} e^{i \mathbf{k} \cdot \mathbf{r}} \int_{\Omega_t} \left[ \rho_{ext}(\mathbf{r}') + \rho_{NL,h}(\mathbf{r}') + \rho_c(\mathbf{r}') \right] e^{-i \mathbf{k} \cdot \mathbf{r}'} dx' dy' \tag{A8}$$

$$\phi_\Delta(\mathbf{r}) = \frac{1}{\varepsilon_h^0} \int_{\Omega_t} \left[ \rho_{NL,h}(x', y', z) + \rho_c(x', y', z) \right] \Phi_{reg}(x - x', y - y') dx' dy' \tag{A9}$$

where

$$\Phi_{reg}(x, y) = \frac{1}{A_c} \sum_{\mathbf{J} \neq (0,0)} \frac{1}{\mathbf{k_J} \cdot \mathbf{k_J} + k_z^2} e^{i \mathbf{k_J} \cdot \mathbf{r}}. \tag{A10}$$

Notice that because $\rho_{ext} = \langle \rho_{ext} \rangle$, the fluctuating potential $\phi_\Delta$ does not depend explicitly on $\rho_{ext}$. It should be clear that Eq. (A9) is only valid for Bloch waves. To make further progress and obtain a formula for $\phi_\Delta$ that is valid for an arbitrary field distribution, we make one additional simplification: we neglect the dependence on $\mathbf{k}$ in Eq.(A10). Moreover, since the fluctuating part of the electric field is given by $\mathbf{e}_\Delta = -\nabla \{\phi_\Delta\}$ it is seen that $\mathbf{e}_{\Delta,z} \equiv \mathbf{e}_\Delta \cdot \hat{\mathbf{z}}$ depends on $\partial_z \rho_c$ and $\partial_z \rho_{NL,h}$ and thus in principle should be small as compared to the transverse components. Hence, we also neglect $\mathbf{e}_{\Delta,z}$. Based on these considerations, we obtain the following explicit formula for $\mathbf{e}_\Delta$

$$\mathbf{e}_\Delta(\mathbf{r}) = -\nabla_t \int_{\Omega_t} \frac{1}{\varepsilon_h^0} \left[ \rho_{NL,h}(x',y',z) + \rho_c(x',y',z) \right] \Phi_{reg,0}(x-x', y-y') dx'dy' \quad \text{(A11)}$$

where $\nabla_t = (\partial_x, \partial_y, 0)$ and,

$$\Phi_{reg,0}(x,y) = \frac{1}{A_c} \sum_{\mathbf{J} \neq (0,0)} \frac{1}{\mathbf{k}_\mathbf{J}^0 \cdot \mathbf{k}_\mathbf{J}^0} e^{i\mathbf{k}_\mathbf{J}^0 \cdot \mathbf{r}} . \quad \text{(A12)}$$

Note that within the considered approximations $\mathbf{e}_\Delta$ only has $x$ and $y$ components, and is written in terms of the polarization charges induced in the metal and in the dielectric, relative to the uniform host background.

## II. The additional potential

We can generalize the theory of Ref. [25], and define the additional potential as the average potential drop from the nanowire in a given cell to the respective cell boundary, so that:

$$\varphi_w = \frac{1}{2\pi} \int_0^{2\pi} \int_{r_w}^{a/2} \mathbf{e} \cdot \hat{\boldsymbol{\rho}} \, d\rho d\theta \approx \frac{1}{2\pi} \int_0^{2\pi} \int_{r_w}^{a/2} \mathbf{e}_\Delta \cdot \hat{\boldsymbol{\rho}} \, d\rho d\theta , \quad \text{(A13)}$$

where $(\rho, \theta, z)$ represent a system of cylindrical coordinates centered at the pertinent nanowire. The second identity follows from the fact that the macroscopic field $\mathbf{E}$ varies slowly in the unit cell, and thus does not contribute to the additional potential. Notice that for the purpose of calculating the additional potential the unit cell is modeled as a circular region with radius $a/2$ (another possibility is to consider a radius $R = a/\sqrt{\pi} = 0.56a$, so that $A_c = a^2 = \pi R^2$). Let us put $\mathbf{e}_\Delta = -\nabla_t \varphi$, where $\varphi$ is the electric potential implicitly defined by Eq. (A11). Then, it is easy to check that $\varphi_w$ can be written as a surface integral over the dielectric region of the transverse unit cell:

$$\varphi_w = \frac{-1}{2\pi} \int \nabla_t \cdot \left( \varphi \frac{\hat{\boldsymbol{\rho}}}{\rho} \right) ds . \tag{A14}$$

We used the fact that $\nabla \cdot \left( \frac{\hat{\boldsymbol{\rho}}}{\rho} \right) = 0$ in the dielectric region of the unit cell. Next, we apply Gauss's theorem to transform the surface integral into two line integrals:

$$\varphi_w = \frac{1}{2\pi r_w} \int_{\partial D} \varphi \, dl + \frac{-1}{2\pi} \int_{\partial \Omega_t} \varphi \frac{1}{\rho} \hat{\boldsymbol{\rho}} \cdot \hat{\mathbf{n}} \, dl . \tag{A15}$$

The first line integral is over the circumference of the nanowire ($\partial D$), and the second line integral is over the boundary of the unit cell ($\partial \Omega_t$); $\hat{\mathbf{n}}$ is the outward unit vector normal to $\partial \Omega_t$. Because $\varphi / \rho$ is expected to be much larger near $\partial D$ than near $\partial \Omega_t$, the contribution from the second line integral can be neglected. Hence, substituting the formula for $\varphi$ [see Eq. (A11)] into the previous equation, we obtain the final result for the value of $\varphi_w$ in the pertinent unit cell ($\Omega_t$):

$$\varphi_w = \int_{\Omega_t} \frac{1}{\varepsilon_h^0} \left[ \rho_{NL,h}(x', y', z) + \rho_c(x', y', z) \right] \Phi_{w0}(x', y') \, dx' dy' \tag{A16}$$

where

$$\begin{aligned}
\Phi_{w0}(x, y) &= \frac{1}{2\pi r_w} \int_{\partial D} \Phi_{reg,0}(x - x', y - y') \, dl \\
&= \frac{1}{A_c} \sum_{\mathbf{J} \neq (0,0)} \frac{1}{\mathbf{k}_\mathbf{J}^0 \cdot \mathbf{k}_\mathbf{J}^0} e^{i \mathbf{k}_\mathbf{J}^0 \cdot \mathbf{r}} J_0 \left( \left| \mathbf{k}_\mathbf{J}^0 \right| r_w \right)
\end{aligned} \tag{A17}$$

being $J_0$ the zero-th order Bessel function of $1^{st}$ kind. For thin nanowires the volumetric charge density $\rho_c$ in the metal can be replaced by an equivalent surface charge density $\sigma_l$ located at the nanowire surface ($\partial D$), so that $\rho_c dx' dy' = \dfrac{\sigma_l}{2\pi r_w} dl'$. Supposing also that $\sigma_l$ is approximately constant over $\partial D$, we find that:

$$\varphi_w = \frac{\sigma_l(z)}{\varepsilon_h^0} \frac{1}{2\pi r_w} \int_{\partial D} \Phi_{w0}(x', y') dl' + \int_{\Omega_t} \frac{1}{\varepsilon_h^0} \rho_{NL,h}(x', y', z) \Phi_{w0}(x', y') dx' dy'. \tag{A18}$$

For future reference, we note that for a dielectric with a linear response $\rho_{NL,h} = 0$. Thus, using $\rho_c dx' dy' = \dfrac{\sigma_l}{2\pi r_w} dl'$ and Eq. (A17), it is seen that the fluctuating part of the electric field [Eq. (A11)] can be written as:

$$\mathbf{e}_\Delta(\mathbf{r})\Big|_{\substack{\text{linear}\\\text{host}}} = -\frac{\sigma_l}{\varepsilon_h^0} \nabla_t \Phi_{w0}. \tag{A19}$$

Hence, the $\Phi_{w0}(x, y)$ can be regarded as the potential that determines the fluctuating part of the electric potential for an array of identical point charges (more correctly, identical linear distributions of charge) placed at the lattice points. The self-field created by the charges in the unit cell gives the dominant contribution to $\mathbf{e}_\Delta$ (notice that the charge in the unit cell does not contribute to $\mathbf{E}$, because the corresponding electric field is radial). Thus, we can estimate that $-\nabla_t \Phi_{w0} \approx \dfrac{1}{2\pi\rho}$. In Ref. [25] the contribution from the nearest nanowire neighbor was also included in the calculation of $\mathbf{e}_\Delta$. One can infer from the theory of Ref. [25], that to a first approximation:

$$-\nabla_t \Phi_{w0} \approx \left[\frac{1}{2\pi\rho} - \frac{1}{2\pi(a-\rho)}\right] \hat{\boldsymbol{\rho}}. \tag{A20}$$

being the second term of the formula the correction that includes the effect of the nearest neighbor.

### III. The relation between $\sigma_l$ and $\varphi_w$

Next, we determine the relation between $\sigma_l$ and $\varphi_w$ under the approximation of a weak nonlinearity in the dielectric response. It is evident from the definition and Eq. (A18) that:

$$\sigma_l = C^0 \varphi_w - \frac{C^0}{\varepsilon_h^0} \int_{\Omega_t} \rho_{NL,h}(x', y', z) \Phi_{w0}(x', y') dx' dy'. \quad (A21)$$

with $C^0$ the p.u.l. capacitance for the linear case, defined so that

$$\frac{1}{C^0} = \frac{1}{\varepsilon_h^0} \frac{1}{2\pi r_w} \int_{\partial D} \Phi_{w0}(x', y') dl'$$
$$= \frac{1}{\varepsilon_h^0} \frac{1}{A_c} \sum_{\mathbf{J} \neq (0,0)} \frac{\left[ J_0\left( \left| \mathbf{k}_\mathbf{J}^0 \right| r_w \right) \right]^2}{\mathbf{k}_\mathbf{J}^0 \cdot \mathbf{k}_\mathbf{J}^0} \quad (A22)$$

The above formula for $C^0$ is different from that derived in Ref. [25] (see Eq. (8)), but for thin wires the two formulas yield very similar numerical results. For example for $r_w/a = 0.01$ Eq. (8) and Eq. (A22) [summing 500×500 terms] predict that $C^0/\varepsilon_h^0$ is 1.95 and 1.91, respectively. Thus, for simplicity and to be consistent with Ref. [25], in this work we adopt Eq. (8) as the definition of $C^0$.

Using $\rho_{NL,h} = \varepsilon_h^0 \nabla \cdot (\delta \varepsilon \nabla \phi) = -\varepsilon_h^0 \nabla \cdot (\delta \varepsilon \mathbf{e})$ in Eq. (A21) it is possible to write after integration by parts:

$$\sigma_l = C^0 \varphi_w - C^0 \int_{\Omega_t} \delta \varepsilon \, \mathbf{e} \cdot \nabla_t \Phi_{w0}(x', y') dx' dy'$$
$$= C^0 \varphi_w - C^0 A_c \langle \delta \varepsilon \, \mathbf{e} \cdot \nabla_t \Phi_{w0} \rangle \quad . \quad (A23)$$

In the second identity, we approximated the surface integral in one cell by the operation of spatial averaging. To make further progress, we consider a weak-nonlinear response so that $\delta \varepsilon \ll 1$. When $\delta \varepsilon = 0$ from Eq. (A19) and Eq. (A23) we know that $-\nabla_t \Phi_{w0} = \dfrac{\varepsilon_h^0}{C^0 \varphi_w} \mathbf{e}_\Delta$. Thus, neglecting corrections of the second order in $\delta\varepsilon$ in Eq. (A23) we obtain Eq. (7).

## Appendix B: The spatial average of relevant products of the microscopic fields

In this Appendix, we calculate the spatial average of several expressions that involve products of the microscopic fields, assuming always that the host dielectric has a *linear* response. We start with the calculation of $\langle \mathbf{e}^* \cdot \mathbf{e} \rangle$. Using the decomposition $\mathbf{e} = \mathbf{E} + \mathbf{e}_\Delta$ [Eq. (3)], and noting that because of Eq. (A11) the fluctuating part of the microscopic electric field $\mathbf{e}_\Delta$ does not have a $z$-component, it is evident that $\mathbf{e} = \mathbf{E} - \nabla_t \varphi$, being $\varphi$ the potential defined implicitly in Eq. (A11) with $\rho_{NL,h} = 0$. Using the identity $\langle f\, \mathbf{F} \rangle = \langle f \rangle \mathbf{F}$, which is valid for a generic functions $f$ and $\mathbf{F}$ when $\mathbf{F}$ is a "macroscopic" field that varies slowly on the scale of the unit cell, we find that:

$$\langle \mathbf{e}^* \cdot \mathbf{e} \rangle = \mathbf{E}^* \cdot \mathbf{E} + \langle |\nabla_t \varphi|^2 \rangle \approx \mathbf{E}^* \cdot \mathbf{E} + \dfrac{1}{A_c} \int |\nabla_t \varphi|^2 ds . \tag{B1}$$

In the second identity, we identify the averaging operator with the spatial averaging in one cell, which is valid for Bloch waves. The integration is taken over the dielectric region because $\mathbf{e}^* \cdot \mathbf{e}$ must be taken equal to zero in the metal region. In case the dielectric has a linear response, we have $\nabla^2 \phi = -\rho_{ext} / \varepsilon_h^0$ in the dielectric region [Eq. (A2)], and hence $\nabla^2 \varphi = 0$. As discussed in Appendix A, the dependence on $z$ of $\varphi$ can be neglected and thus we can assume that $\nabla_t^2 \varphi = 0$ in the dielectric region, so that $|\nabla_t \varphi|^2 = \nabla_t \cdot (\varphi^* \nabla_t \varphi)$. We use this result to integrate Eq. (B1) by parts, and transform the surface integral into a line integral over the boundary of the cell and a line integral over the boundary of the nanowire ($\partial D$). The former line integral vanishes for Bloch waves. Thus, we obtain:

$$\langle \mathbf{e}^* \cdot \mathbf{e} \rangle \approx \mathbf{E}^* \cdot \mathbf{E} + \frac{1}{A_c} \int_{\partial D} \varphi^* \left( -\frac{\partial \varphi}{\partial n} \right) dl, \tag{B2}$$

where $\frac{\partial \varphi}{\partial n}$ stands for the normal derivative. Noting that the equivalent surface charge density in the nanowire satisfies in the vicinity of $\partial D$ $\frac{\sigma_l}{2\pi r_w \varepsilon_h^0} = -\frac{\partial \phi}{\partial n} \approx -\frac{\partial \varphi}{\partial n}$, we can write

$$\langle \mathbf{e}^* \cdot \mathbf{e} \rangle \approx \mathbf{E}^* \cdot \mathbf{E} + \frac{1}{A_c} \varphi_w^* \frac{\sigma_l}{\varepsilon_h^0}. \tag{B3}$$

We used the property $\varphi_w \approx \frac{1}{2\pi r_w} \int_{\partial D} \varphi dl$ (see Eq. (A15)). Finally, we use Eq. (A23) with $\delta \varepsilon = 0$ and $C^0 = A_c \varepsilon_h^0 \beta_p^2$ to obtain:

$$\langle \mathbf{e} \cdot \mathbf{e}^* \rangle \approx \mathbf{E} \cdot \mathbf{E}^* + \beta_p^2 |\varphi_w|^2. \tag{B4}$$

Even though, strictly speaking, this derivation only holds in case of microscopic fields with a Bloch-type spatial variation, it will be assumed that it remains valid in case of arbitrary fields. It should be clear that derived result also implies that

$$\langle \mathbf{e}_\Delta \cdot \mathbf{e}_\Delta^* \rangle = \beta_p^2 |\varphi_w|^2, \tag{B5}$$

where $\mathbf{e}_\Delta$ is the fluctuating part of the microscopic electric field. From Eq. (A19) and the fact that $\Phi_{w0}$ is real valued, we can also conclude that $\langle \mathbf{e}_\Delta \cdot \mathbf{e}_\Delta \rangle = \langle \mathbf{e}_\Delta \cdot \mathbf{e}_\Delta^* \rangle \varphi_w / \varphi_w^*$, or equivalently:

$$\langle \mathbf{e}_\Delta \cdot \mathbf{e}_\Delta \rangle = \beta_p^2 \varphi_w^2. \tag{B6}$$

It is interesting to mention that Eq. (B4) can also be derived based on the formula for the electromagnetic energy density. Specifically, it was shown in Ref. [39] (see Eq. (62) of Ref. [39] generalized to the case of time-harmonic fields) that the electromagnetic energy density can be written in terms of the macroscopic fields as follows

$$W_{mac} = \frac{1}{4}\varepsilon_h^0 \mathbf{E}\cdot\mathbf{E}^* + \frac{1}{4}\mu_0 \mathbf{H}\cdot\mathbf{H}^* + \frac{1}{4A_c}\left(L_{tot}|I|^2 + C^0|\varphi_w|^2\right), \quad \text{where} \quad L_{tot} = L + L_{kin} \quad \text{and the}$$

current $I$ are defined as in Sect. II. Because the additional potential is created by the charges in the metal, it is possible to identify $W_{E,mac} = \frac{1}{4}\varepsilon_h^0 \mathbf{E}\cdot\mathbf{E}^* + \frac{1}{4A_c}C^0|\varphi_w|^2$ as the electric energy density. On the other hand, from a microscopic point of view, the electric energy density is $W_{E,mic} = \frac{1}{4}\frac{\partial}{\partial \omega}(\varepsilon'\omega)\mathbf{e}\cdot\mathbf{e}^*$. Thus, it is reasonable to expect that $W_{E,mac} = \langle W_{E,mic} \rangle$. Since for good conductors the energy stored inside the metal is in principle negligible, one can write $\langle W_{E,mic} \rangle \approx \frac{1}{4}\varepsilon_h^0 \langle \mathbf{e}\cdot\mathbf{e}^* \rangle$ and hence the relation $W_{E,mac} = \langle W_{E,mic} \rangle$ is equivalent Eq. (B4).

Next, we compute $\langle \mathbf{e}^* \cdot \mathbf{e}\,\mathbf{e}_\Delta \rangle$ in terms of the macroscopic fields. Using the decomposition $\mathbf{e} = \mathbf{E} + \mathbf{e}_\Delta$ [Eq. (3)], the result $\langle \mathbf{e}_\Delta \rangle = 0$, and the fact that $\mathbf{E}$ can be brought out of the averaging operator because it is a slowly varying macroscopic function, it is easily found that:

$$\langle \mathbf{e}^* \cdot \mathbf{e}\,\mathbf{e}_\Delta \rangle = \langle \mathbf{e}_\Delta \mathbf{e}_\Delta \rangle \cdot \mathbf{E}^* + \langle \mathbf{e}_\Delta \mathbf{e}_\Delta^* \rangle \cdot \mathbf{E} + \langle \mathbf{e}_\Delta \cdot \mathbf{e}_\Delta^* \mathbf{e}_\Delta \rangle. \tag{B7}$$

In case of dielectric host with a linear response the fluctuating part of the electric field satisfies Eq. (A19). The function $\Phi_{w0}(x,y)$ is given by Eq. (A17), and has the following symmetries in case of a square lattice:

$$\Phi_{w0}(x,y) = \Phi_{w0}(\pm x, y) = \Phi_{w0}(x, \pm y) = \Phi_{w0}(y,x). \tag{B8}$$

This shows that $\mathbf{e}_\Delta \cdot \mathbf{e}_\Delta^*$ is an even function of both $x$ and $y$, whereas $\mathbf{e}_{\Delta,x} \equiv \mathbf{e}_\Delta \cdot \hat{\mathbf{x}}$ ($\mathbf{e}_{\Delta,y} \equiv \mathbf{e}_\Delta \cdot \hat{\mathbf{y}}$) is an odd function of $x$ ($y$). Therefore, if one identifies the averaging operator with the surface integral in one cell, it is readily seen that:

$$\langle \mathbf{e}_\Delta \cdot \mathbf{e}_\Delta^* \mathbf{e}_\Delta \rangle = 0 = \langle \mathbf{e}_\Delta \cdot \mathbf{e}_\Delta \mathbf{e}_\Delta \rangle. \tag{B9}$$

Using similar arguments, one can see that $\langle \mathbf{e}_\Delta \mathbf{e}_\Delta \rangle = \langle \mathbf{e}_{\Delta,x} \mathbf{e}_{\Delta,x} \rangle \hat{\mathbf{x}}\hat{\mathbf{x}} + \langle \mathbf{e}_{\Delta,y} \mathbf{e}_{\Delta,y} \rangle \hat{\mathbf{y}}\hat{\mathbf{y}}$ and, analogously, that $\langle \mathbf{e}_\Delta \mathbf{e}_\Delta^* \rangle = \langle \mathbf{e}_{\Delta,x} \mathbf{e}_{\Delta,x}^* \rangle \hat{\mathbf{x}}\hat{\mathbf{x}} + \langle \mathbf{e}_{\Delta,y} \mathbf{e}_{\Delta,y}^* \rangle \hat{\mathbf{y}}\hat{\mathbf{y}}$. Obviously, for a square lattice $\langle \mathbf{e}_{\Delta,x} \mathbf{e}_{\Delta,x} \rangle = \langle \mathbf{e}_{\Delta,y} \mathbf{e}_{\Delta,y} \rangle$. On the other hand, because $\Phi_{w0}$ is real-valued and from Eq. (A19) and $\sigma_l = C^0 \varphi_w$, we can write $\langle \mathbf{e}_{\Delta,x} \mathbf{e}_{\Delta,x} \rangle = \langle \mathbf{e}_{\Delta,x} \mathbf{e}_{\Delta,x}^* \rangle \frac{\varphi_w}{\varphi_w^*} = \frac{1}{2} \langle \mathbf{e}_\Delta \cdot \mathbf{e}_\Delta^* \rangle \frac{\varphi_w}{\varphi_w^*}$. Substituting the previous results into Eq. (B7), and using Eq. (B5), we finally obtain that:

$$\langle \mathbf{e}^* \cdot \mathbf{e} \mathbf{e}_\Delta \rangle = \frac{1}{2} \beta_p^2 \varphi_w^2 \mathbf{E}_t^* + \frac{1}{2} \beta_p^2 |\varphi_w|^2 \mathbf{E}_t, \tag{B10}$$

where $\mathbf{E}_t = \bar{\bar{\mathbf{I}}}_t \cdot \mathbf{E}$ is the transverse part of the electric field, with $\bar{\bar{\mathbf{I}}}_t = \hat{\mathbf{x}}\hat{\mathbf{x}} + \hat{\mathbf{y}}\hat{\mathbf{y}}$.

To conclude this Appendix, we determine $\langle |\mathbf{e}_\Delta|^2 \mathbf{e}_\Delta \cdot \mathbf{e}_\Delta \rangle$. To this end, we identify again the averaging operator with the integral in the dielectric region. Using Eq. (A19) and $\sigma_l = C^0 \varphi_w$, it is found that:

$$\left\langle \left|\mathbf{e}_{\Delta}\right|^{2} \mathbf{e}_{\Delta} \cdot \mathbf{e}_{\Delta} \right\rangle = \varphi_{w}^{*}\varphi_{w}^{3} \left(\frac{C^{0}}{\varepsilon_{h}^{0}}\right)^{4} \frac{1}{A_{c}} \int_{\text{dielectric}} \left|\nabla_{t}\Phi_{w0}\right|^{2} \nabla_{t}\Phi_{w0} \cdot \nabla_{t}\Phi_{w0}\, dxdy\,. \quad \text{(B11)}$$

Because $\Phi_{w0}$ is real-valued and using $\beta_{p}^{2} = C^{0}/\left(\varepsilon_{h}^{0}A_{c}\right)$, the previous formula can be rewritten as:

$$\left\langle \left|\mathbf{e}_{\Delta}\right|^{2} \mathbf{e}_{\Delta} \cdot \mathbf{e}_{\Delta} \right\rangle = \tilde{B}\, \beta_{p}^{4} \varphi_{w}^{*}\varphi_{w}^{3}, \quad \text{(B12)}$$

with the dimensionless parameter $\tilde{B}$ defined by,

$$\tilde{B} = \beta_{p}^{4} A_{c}^{3} \int_{0}^{2\pi} d\theta \int_{r_{w}}^{a/2} d\rho\, \rho \left|\nabla_{t}\Phi_{w0}\right|^{4}. \quad \text{(B13)}$$

The unit cell was approximated by a circular region with radius $a/2$. The parameter $\tilde{B}$ can be numerically evaluated by substituting Eq. (A17) into the previous formula. Alternatively, we can use the approximation (A20) to write:

$$\tilde{B} = 2\pi\, \beta_{p}^{4} A_{c}^{3} \int_{r_{w}}^{a/2} \left(\frac{1}{2\pi\rho} - \frac{1}{2\pi(a-\rho)}\right)^{4} \rho\, d\rho\,. \quad \text{(B14)}$$

This integral can be calculated explicitly, but the final expression is too long to show here.

## References:

[1] P. A. Belov, Y. Hao, and S. Sudhakaran, *Phys. Rev. B*, **73**, 033108, (2006).

[2] P. Ikonen, C. Simovski, S. Tretyakov, P. Belov, and Y. Hao, *Appl. Phys. Lett.*, **91**, 104102 (2007).

[3] G. Shvets, S. Trendafilov, J. B. Pendry, and A. Sarychev, *Phys. Rev. Lett.*, **99**, 053903 (2007).

[4] I. V. Lindell, S. A. Tretyakov, K. I. Nikoskinen, S. Ilvonen, *Microwave Opt. Tech. Lett.* **31**, 129 (2001).

[5] D. R. Smith and D. Schurig, *Phys. Rev. Lett.* **90**, 077405 (2003).

[6] Y. Liu, G. Bartal, X. Zhang, *Opt. Express* **16**, 15439 (2008).

[7] J. Yao, Z. Liu, Y. Liu, Y. Wang, C. Sun, G. Bartal, A. M. Stacy, X. Zhang, *Science* **321**, 930 (2008).

[8] M. A. Noginov, H. Li, Y. A. Barnakov, D. Dryden, G. Nataraj, G. Zhu, C. E. Bonner, M. Mayy, Z. Jacob, and E. E. Narimanov, *Opt. Lett.* **35**, 1863 (2010).

[9] A. N. Poddubny, P. A. Belov, Y. S. Kivshar, *Phys. Rev. A* **84**, 023807 (2011).

[10] Z. Jacob; I. I. Smolyaninov, E. E. Narimanov, *Appl. Phys. Lett..* **100**, 181105 (2012); H. N. S. Krishnamoorthy; Z. Jacob, E. Narimanov, I. Kretzschma, V. M. Menon, *Science*, **336**, 205 (2012).

[11] D. E. Fernandes, S. I. Maslovski, M. G. Silveirinha, *Phys. Rev. B* **85**, 155107 (2012).

[12] V. V. Vorobev and A. V. Tyukhtin, *Phys. Rev. Lett.* **108,** 184801 (2012).

[13] I. S. Nefedov and C. R. Simovski, *Phys. Rev. B* **84**, 195459 (2011).

[14] S. I. Maslovski, M. G. Silveirinha, *Phys. Rev. A* **82**, 022511 (2010).

[15] S. I. Maslovski, M. G. Silveirinha, *Phys. Rev. A* **83**, 022508 (2011).

[16] F. Ye, D. Mihalache, B. Hu and N. C. Panoiu, *Phys. Rev. Lett.* **104**, 106802 (2010).

[17] Y. Liu, G. Bartal, D. A. Genov, and X. Zhang, *Phys. Rev. Lett.* **99**, 153901 (2007).

[18] D. N. Christodoulides, R. I. Joseph, *Opt. Lett.*, **13**, 794, (1988).

[19] H. S. Eisenberg, Y. Silberberg, R. Morandotti, A. R. Boyd, and J. S. Aitchison, *Phys. Rev. Lett.*, **81**, 3383 (1998).

[20] F. Ye, D. Mihalache, B. Hu, and N. C. Panoiu, *Opt. Lett.*, **36**, 1179, (2011).

[21] Y. Kou, F. Ye and X. Chen, *Phys. Rev. A*, **84**, 033855 (2011).

[22] J.-Y. Yan, L. Li, and J. Xiao, *Optics Express,* **20**, 1945, (2012)

[23] P. A. Belov, R. Marqués, S. I. Maslovski, I.S. Nefedov, M. Silveirinha, C. R. Simovski, S. A. Tretyakov, *Phys. Rev. B*, **67**, 113103 (2003).

[24] M. G. Silveirinha, *Phys. Rev. E*, **73**, 046612, (2006).

[25] S. I. Maslovski, M. G. Silveirinha, *Phys. Rev. B* **80**, 245101, (2009).

[26] A. A. Zharov, I. V. Shadrivov, Y. S. Kivshar, *Phys. Rev. Lett.* **91**, 037401 (2003).

[27] M. Lapine, M. Gorkunov, K. H. Ringhofer, *Phys. Rev. E*, **67**, 065601(R), (2003).

[28] Y. Zeng, D. A. R. Dalvit, J. O'Hara, S. A. Trugman, *Phys. Rev. B* **85**, 125107 (2012).

[29] V. M. Agranovich, Y. R. Shen, R. H. Baughman, and A. A. Zakhidov, *Phys. Rev. B* **69**, 165112 (2004).

[30] D. A. Martin and M. Hoyuelos, *Phys. Rev. A* **82**, 033841 (2010).

[31] A. Ciattoni and E. Spinozzi, *Phys. Rev. A* **85**, 043806 (2012).

[32] R. W. Boyd, *Nonlinear Optics* (3$^{rd}$ Ed.), *Academic Press*, (San Diego, CA, USA) 2008.

[33] G. Russakov, *Am. J. Physics*, **38**, 1188, (1970).

[34] J. D. Jackson, *Classical Electrodynamics* (Wiley, 1998).

[35] V. Agranovich and V. Ginzburg, *Spatial Dispersion in Crystal Optics and the Theory of Excitons*. New York: Wiley- Interscience, 1966.

[36] M. A. Ordal, R. J. Bell, R. W. Alexander, Jr., L. L. Long, and M. R. Querry, *Appl. Opt*. **24**, 4493, (1985).

[37] M. G. Silveirinha, P. A. Belov, C. R. Simovski, *Phys. Rev. B* **75**, 035108 (2007).

[38] M. G. Silveirinha, P. A. Belov, C. R. Simovski, *Opt. Lett.*, **33**, 1726, (2008).

[39] M. G. Silveirinha, S. I. Maslovski, *Phys. Rev. B*, **85**, 155125 (2012).

[40] Q. Lin, J. Zhang, G. Piredda, R. W. Boyd, P. M. Fauchet, G. P. Agrawal, *Appl. Phys. Lett.* **91**, 021111 (2007)

[41] M. G. Silveirinha, C. A. Fernandes, *IEEE Trans. on Antennas and Propag.*, **53**, 347, (2005).-43-